\documentclass[runningheads]{llncs}
\usepackage[T1]{fontenc}
\usepackage{cite}
\usepackage{amsmath,amssymb,amsfonts}
\usepackage{algorithmic}
\usepackage{graphicx,color}
\usepackage{textcomp}
\usepackage{xcolor}
\usepackage{hyperref}
\hypersetup{hidelinks=true}
\usepackage{algorithm,algorithmic}
\usepackage{acronym}
\usepackage{pgfplots}
\usepackage{microtype}
\def\BibTeX{{\rm B\kern-.05em{\sc i\kern-.025em b}\kern-.08em
    T\kern-.1667em\lower.7ex\hbox{E}\kern-.125emX}}
\AtBeginDocument{\definecolor{tmlcncolor}{cmyk}{0.93,0.59,0.15,0.02}\definecolor{NavyBlue}{RGB}{0,86,125}}

\acrodef{avrmccc}[A-VRMCCC]{augmented Volterra recursive maximum complex correntropy criterion}
\acrodef{bc}[BC]{bias-compensated}
\acrodef{bl}[BL]{bilinear}
\acrodef{cv}[CV]{complex-valued}
\acrodef{fir}[FIR]{finite impulse response}
\acrodef{iq}[IQ]{in-phase and quadrature-phase}
\acrodef{ise}[ISE]{instantaneous squared error}
\acrodef{lut}[LUT]{lookup table}
\acrodef{nm}[NM]{nomalized misalignment}
\acrodef{mmse}[MMSE]{minimum mean squared error}
\acrodef{mse}[MSE]{mean squared error}
\acrodef{miso}[MISO]{multiple-input-single-output}
\acrodef{nn}[NN]{neural network}
\acrodef{ofdm}[OFDM]{orthogonal frequency-division multiplexing}
\acrodef{pa}[PA]{power amplifier}
\acrodef{rv}[RV]{real-valued}
\acrodef{siso}[SISO]{single-input-single-output}
\acrodef{tls}[TLS]{total least squares}
\acrodef{wss}[WSS]{wide sense stationary}
\acrodef{sota}[SOTA]{state-of-the-art}

\acrodef{kf}[KF]{Kalman filter}
\acrodef{aclkf1}[AC-LKF]{augmented complex-valued linear Kalman filter}
\acrodef{aclkf2}[AC-LKF]{augmented complex-valued linear Kalman filter}
\acrodef{aclkf}[AC-LKF]{augmented complex-valued linear \acs{kf}}

\acrodef{ekf1}[EKF]{extended Kalman filter}
\acrodef{ekf}[EKF]{extended \acs{kf}}
\acrodef{acekf1}[AC-EKF]{augmented complex-valued extended Kalman filter}
\acrodef{acekf2}[AC-EKF]{augmented complex-valued extended Kalman filter}
\acrodef{acekf}[AC-EKF]{augmented complex-valued extended \acs{kf}}

\acrodef{lms}[LMS]{least mean squares}
\acrodef{ls}[LS]{least squares}

\acrodef{nlms1}[NLMS]{normalized least mean squares}
\acrodef{nlms}[NLMS]{normalized \ac{lms}}

\acrodef{rls}[RLS]{recursive least squares}

\acrodef{wf}[WF]{Wiener filter}

\acrodef{2rblms1}[$2\mathbb{R}$-BLMS]{$2\mathbb{R}$ bilinear least mean squares}
\acrodef{2rblms2}[$2\mathbb{R}$-BLMS]{$2\mathbb{R}$ bilinear least mean squares}
\acrodef{2rblms}[$2\mathbb{R}$-BLMS]{$2\mathbb{R}$ bilinear \acs{lms}}

\acrodef{2rbnlms1}[$2\mathbb{R}$-BNLMS]{$2\mathbb{R}$ bilinear normalized least mean squares}
\acrodef{2rbnlms2}[$2\mathbb{R}$-BNLMS]{$2\mathbb{R}$ bilinear normalized least mean squares}
\acrodef{2rbnlms3}[$2\mathbb{R}$-BNLMS]{$2\mathbb{R}$ bilinear normalized \acs{lms}}
\acrodef{2rbnlms}[$2\mathbb{R}$-BNLMS]{$2\mathbb{R}$ bilinear \acs{nlms}}

\acrodef{4rblms1}[$4\mathbb{R}$-BLMS]{$4\mathbb{R}$ bilinear least mean squares}
\acrodef{4rblms2}[$4\mathbb{R}$-BLMS]{$4\mathbb{R}$ bilinear least mean squares}
\acrodef{4rblms}[$4\mathbb{R}$-BLMS]{$4\mathbb{R}$ bilinear \acs{lms}}

\acrodef{4rbnlms1}[$4\mathbb{R}$-BNLMS]{$4\mathbb{R}$ bilinear normalized least mean squares}
\acrodef{4rbnlms2}[$4\mathbb{R}$-BNLMS]{$4\mathbb{R}$ bilinear normalized least mean squares}
\acrodef{4rbnlms3}[$4\mathbb{R}$-BNLMS]{$4\mathbb{R}$ bilinear normalized \acs{lms}}
\acrodef{4rbnlms}[$4\mathbb{R}$-BNLMS]{$4\mathbb{R}$ bilinear \acs{nlms}}

\acrodef{bkf1}[BKF]{bilinear Kalman filter}
\acrodef{bkf2}[BKF]{bilinear Kalman filter}
\acrodef{bkf}[BKF]{bilinear \acs{kf}}
\acrodef{acbkf1}[AC-BKF]{augmented complex-valued bilinear Kalman filter}
\acrodef{acbkf2}[AC-BKF]{augmented complex-valued bilinear Kalman filter}
\acrodef{acbkf3}[AC-BKF]{augmented complex-valued bilinear Kalman filter}
\acrodef{acbkf4}[AC-BKF]{augmented complex-valued bilinear \acs{kf}}
\acrodef{acbkf}[AC-BKF]{augmented complex-valued \acs{bkf}}

\acrodef{cblms1}[C-BLMS]{complex-valued bilinear least mean squares}
\acrodef{cblms2}[C-BLMS]{complex-valued bilinear least mean squares}
\acrodef{cblms3}[C-BLMS]{complex-valued bilinear least mean squares}
\acrodef{cblms}[C-BLMS]{complex-valued bilinear least mean squares}
\acrodef{cllms}[C-LLMS]{complex-valued linear \acs{lms}}

\acrodef{cbls1}[C-BLS]{complex-valued bilinear least squares}
\acrodef{cbls2}[C-BLS]{complex-valued bilinear least squares}
\acrodef{cbls3}[C-BLS]{complex-valued bilinear least squares}
\acrodef{cbls}[C-BLS]{complex-valued bilinear \acs{ls}}

\acrodef{cbnlms1}[C-BNLMS]{complex-valued bilinear normalized least mean squares}
\acrodef{cbnlms2}[C-BNLMS]{complex-valued bilinear normalized least mean squares}
\acrodef{cbnlms3}[C-BNLMS]{complex-valued bilinear normalized least mean squares}
\acrodef{cbnlms}[C-BNLMS]{complex-valued bilinear normalized \acs{lms}}

\acrodef{cbrls1}[C-BRLS]{complex-valued bilinear recursive least squares}
\acrodef{cbrls2}[C-BRLS]{complex-valued bilinear recursive least squares}
\acrodef{cbrls3}[C-BRLS]{complex-valued bilinear recursive least squares}
\acrodef{cbrls}[C-BRLS]{complex-valued bilinear \acs{rls}}

\acrodef{cbwf1}[C-BWF]{complex-valued bilinear Wiener filter}
\acrodef{cbwf2}[C-BWF]{complex-valued bilinear Wiener filter}
\acrodef{cbwf3}[C-BWF]{complex-valued bilinear Wiener filter}
\acrodef{cbwf}[C-BWF]{complex-valued bilinear Wiener filter}

\acrodef{crblms1}[CR-BLMS]{mixed complex-valued-real-valued bilinear least mean squares}
\acrodef{crblms}[CR-BLMS]{mixed complex-valued-\acs{rv} bilinear \acs{lms}}

\acrodef{crbls1}[CR-BLS]{mixed complex-valued-real-valued bilinear least squares}
\acrodef{crbls}[CR-BLS]{mixed complex-valued-\acs{rv} bilinear \acs{ls}}

\acrodef{crbnlms1}[CR-BNLMS]{mixed complex-valued-real-valued bilinear normalized least mean squares}
\acrodef{crbnlms}[CR-BNLMS]{mixed complex-valued-\acs{rv} bilinear \acs{nlms}}

\acrodef{crbrls1}[CR-BRLS]{mixed complex-valued-real-valued bilinear recursive least squares}
\acrodef{crbrls}[CR-BRLS]{mixed complex-valued-\acs{rv} bilinear \acs{rls}}

\acrodef{crbwf1}[CR-BWF]{mixed complex-valued-real-valued bilinear Wiener filter}
\acrodef{crbwf}[CR-BWF]{mixed complex-valued-\acs{rv} bilinear \acs{wf}}

\acrodef{hsaf1}[HSAF]{Hammerstein spline adaptive filter}
\acrodef{hsaf}[HSAF]{Hammerstein \ac{saf}}
\acrodef{saf}[SAF]{spline adaptive filter}
\acrodef{wsaf1}[WSAF]{Wiener spline adaptive filter}
\acrodef{wsaf}[WSAF]{Wiener \ac{saf}}

\acrodef{wlceilms}[WLC-CEILMS]{widely linear complex-valued estimated-input least mean squares}
\acrodef{lceilms}[LC-CEILMS]{linear complex-valued estimated-input least mean squares}
\acrodef{wlceimccc}[WLC-EIMCCC]{widely linear complex-valued estimated-input maximum complex correntropy criterion}
\acrodef{wlceimcse}[WLC-EIMCSE]{widely linear complex-valued estimated-input minimum complex Shannon entropy}
\acrodef{acgdtls}[ACGDTLS]{augmented complex-valued gradient-descent total least-squares}
\acrodef{vmmtcc}[VMMTCC]{variable maximum mixture complex correntropy}

\newcommand{\ve}{\mathbf}
\newcommand{\m}{\mathbf}

\begin{document}

\title{Bias-Compensated Complex-Valued Bilinear Filtering}
%
%
\author{Bernhard Plaimer\inst{1,2} \and
Andreas Meingassner-Lang\inst{1,2} \and
Yannis Kowalewski\inst{2} \and  
Matthias Wagner\inst{2} \and  
Oliver Lang\inst{2} \and 
Mario Huemer\inst{2}}

\authorrunning{B. Plaimer, A. Meingassner-Lang et al.}

\institute{The authors contributed equally to this work. \and
Institute of Signal Processing, Johannes Kepler University Linz, Austria
\email{Bernhard.Plaimer@jku.at, Andreas.Meingassner-Lang@jku.at}}
%
%
%
\maketitle              
\begin{abstract}
Complex-valued optimal and adaptive filters are widely used to identify unknown systems in a broad range of practical applications. Recently, a number of complex-valued bilinear filters, such as the \ac{cbwf}, and the \ac{cblms} filter, have been proposed to model and identify complex-valued systems, which are bilinear with respect to their coefficients. However, if the input signals are contaminated with noise, the performance of these complex-valued bilinear filters degrades significantly due to an additional noise-induced bias. 

To overcome this issue, we propose novel bias-compensated complex-valued bilinear filters that estimate and account for input-noise statistics. Specifically, a bias-compensated \ac{cbwf} and a bias-compensated \ac{cblms} filter are derived in this work. We further include a convergence analysis for the latter. Since practical applications require knowledge of the input-noise variance, we briefly present an adapted method from literature to estimate this quantity. Finally, simulation results demonstrate the superior performance of the proposed filters compared to their standard (non-bias-compensated) counterparts and several state-of-the-art methods.

\end{abstract}
\section{INTRODUCTION}

The task of system identification is fundamental in digital signal processing and is typically addressed with optimum and adaptive filtering techniques \cite{Diniz_2008_1,Kay_1993_1,Haykin_2014_1}. System identification underlies a broad range of applications, including channel estimation in wireless communication systems, echo-path identification for acoustic echo cancellation, plant modeling for control, and channel modeling in radar. In its most basic and widely studied form, system identification assumes that the input signal is given without any errors. In practice, e.g., measurement and quantization noise may corrupt the input signal, and, as established in the linear filtering literature, this yields biased filter coefficients \cite{Davila_1994_1}. 

To compensate this bias, many optimum and adaptive filters were proposed \cite{Davila_1994_1,Javed_2014_1,Arablouei_2014_1,Yuan_2025_1,Jia_2001_1,Kang_2013_1,Abdolee_2016_1,SungEun_2005_1,douglas1997posteriori}. The filters \cite{Davila_1994_1,Javed_2014_1,Arablouei_2014_1}, which rely on a \ac{tls} approach \cite{Golub_1980_1}, remove the bias implicitly by solving an eigenvalue problem on an augmented data matrix. While well motivated, the required singular value decomposition is computationally costly, which limits the practical use of these methods in real-time adaptive settings. The remaining filters \cite{Yuan_2025_1,Jia_2001_1,Kang_2013_1,Abdolee_2016_1,SungEun_2005_1,douglas1997posteriori} avoid this by minimizing an adapted cost function, or by applying an a posteriori correction term based on an estimation of the bias.

Bias compensation has also been extended beyond real-valued linear systems in two ways. On the one hand, bias-compensated kernel \acs{lms} algorithms \cite{Chien_2025_1,Chien_2025_2} and a general nonlinear \ac{tls} framework \cite{Boggs_1987_1} address systems that have nonlinearities with respect to the input signal. These methods, however, do not cover systems that are nonlinear with respect to their coefficients, such as the bilinear systems considered in \cite{Benesty_2021_1}. On the other hand, several works have extended bias compensation to complex-valued filtering \cite{Zhang_2019_1,Zhang_2025,Dong_2020_1,Cai_2025_1,Huang_2022_1}. The \ac{wlceilms} filter and its linear variant the \ac{lceilms} filter from \cite{Zhang_2019_1} enable unbiased system identification by incorporating denoised input estimates in the update equation, handling both circular and noncircular complex-valued signals. In \cite{Zhang_2025}, the \ac{acgdtls} filter employs a Rayleigh quotient cost function with augmented complex statistics, providing robust performance improvements across varying step-sizes and noise conditions. The methods proposed in \cite{Dong_2020_1,Cai_2025_1} use a correntropy-based cost function to achieve robustness against non-Gaussian noise and impulsive noises while preserving unbiasedness. Similarly, in \cite{Huang_2022_1}, the \ac{wlceimcse} filter combines bias compensation with a Shannon-entropy-based criterion for robustness against non-Gaussian noise. These complex-valued extensions of the real-valued case, however, remain restricted to linear or widely linear systems \cite{Schreier_2010_1}. Consequently, none of the aforementioned bias-compensated filters, real-valued or complex-valued, directly cover systems that are complex-valued and bilinear with respect to their coefficients.

Such complex-valued bilinear systems, however, occur naturally in a variety of applications, including the following two examples. In \ac{iq} imbalance compensation for joint communication and sensing transceivers, which can be cast into a complex-valued bilinear problem, the reference signal used to estimate and compensate transmitter and receiver \ac{iq} imbalance is itself affected by hardware imperfections and measurement noise, so the identification has to be performed from a noisy input signal \cite{Meingassner_2026_1}. Similarly, in automotive radar systems, online calibration of channel imbalances relies on input signals that are themselves noisy, again turning the channel-imbalance problem into a complex-valued bilinear system identification task with a noisy input \cite{Ghafi_2026_1}. 

For the class of complex-valuled bilinear systems, \cite{Plaimer_2025_1} introduced several optimum and adaptive complex-valued bilinear filters, including the \ac{cbwf} and the \ac{cblms} filter. Crucially, \cite{Plaimer_2025_1} assumes noise-free input data. As soon as the input is corrupted by noise an additional, input-noise-induced bias arises. As will be shown later, every bilinear system can be reformulated as a linear one. Consequently, the algorithms in 
\cite{Zhang_2019_1,Zhang_2025,Huang_2022_1,Dong_2020_1,Cai_2025_1} can be applied to complex-valued bilinear systems, albeit at the cost 
of increased number of coefficients and therefore slower convergence. Nevertheless, methods \cite{Zhang_2019_1,Zhang_2025,Dong_2020_1,Cai_2025_1} are included in the simulations as state-of-the-art methods\footnote{\cite{Huang_2022_1} is not included as the method requires additional specification for direct implementation.}. The bias-compensated complex-valued bilinear filters developed in this work resolve this additional bias without increasing the number of coefficients. To the best of the authors' knowledge, no comparable work has been published so far.

This work addresses this open problem. Specifically, we
\begin{itemize}
   \item formalize the problem of biased filter coefficients in complex-valued bilinear filters caused by input noise.
   \item derive a bias-compensated \ac{cbwf} and a bias-compensated \ac{cblms} filter by minimizing an adapted cost function.
   \item provide a convergence analysis of the bias-compensated \ac{cblms} filter.
   \item briefly present an adapted method of \cite{SungEun_2005_1} to estimate the input noise variance.
\end{itemize}
Simulation results, obtained by identifying a complex-valued \ac{miso} system (which can be written as a complex-valued bilinear model) from noisy input data, confirm that the proposed bias-compensated filters substantially outperform their standard (non-bias-compensated) counterparts and several state-of-the-art methods \cite{Zhang_2019_1,Zhang_2025,Dong_2020_1,Cai_2025_1}.


\noindent\textit{Notation and Definitions:}

Lowercase and uppercase boldface letters denote vectors and matrices, respectively. $\left(\cdot\right)^T$ indicates the transposition, and $\left(\cdot\right)^H$ indicates the complex conjugate transposition of a vector or a matrix. Furthermore, $\left(\cdot\right)^*$ indicates the complex conjugation. $\m{I}^{M}$ represents an $M \times M$ identity matrix. The expectation operator is denoted by $\operatorname{E}\left[\cdot\right]$, and the real and imaginary parts of a variable are indicated by $\operatorname{Re}\left[\cdot\right]$ and $\operatorname{Im}\left[\cdot\right]$, respectively. Furthermore, the imaginary unit is represented by $\text{j}$. The vectorization operator is defined as $\widetilde{\ve{a}} = \operatorname{vec}\left[\m{A}\right] = \operatorname{vec}\left[\begin{bmatrix}
	\ve{a}_1	&	\cdots	&	\ve{a}_M
\end{bmatrix}\right] = \begin{bmatrix}
	\ve{a}_1^T & \cdots & \ve{a}_M^T
\end{bmatrix}^T \in \mathbb{C}^{LM}$,
where $\ve{a}_m \in \mathbb{C}^{L}$ denotes the $m$th column of $\m{A} \in \mathbb{C}^{L \times M}$. The operator $\otimes$ is used for the Kronecker product.

\section{Complex-Valued Bilinear Model}\label{sec:complex_valued_biinear_model}
\begin{figure}[!t]
	\centering
	\includegraphics{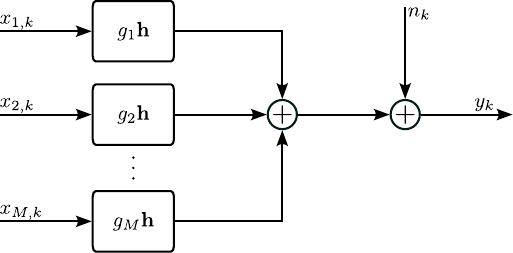}
	\caption{Block diagram of a complex-valued \ac{miso} system.}
	\label{fig:MISO}
\end{figure}

A general complex-valued bilinear model with coefficient vectors $\ve{h} \in \mathbb{C}^L$ and $\ve{g} \in \mathbb{C}^M$ can be described as
\begin{align}
   y_k = \ve{h}^H \m{X}_k \ve{g} + n_k\text{,} \label{eq:bilinear-model}
\end{align}
where $y_k \in \mathbb{C}$ is the output signal, $\m{X}_k \in \mathbb{C}^{L \times M}$ is the input signal matrix, and $n_k \in \mathbb{C}$ is the noise signal at time instance $k$. The bilinear model in \eqref{eq:bilinear-model} may cover a variety of applications. An illustrative complex-valued \ac{miso} system is given in \autoref{fig:MISO}.
In this setting, $\ve{h}$ is a complex-valued channel (modeled as a complex-valued \ac{fir} filter), and $\ve{g} = \begin{bmatrix}
   g_1 & g_2 &  \cdots   &  g_M
\end{bmatrix}^T$ contains complex-valued per-channel scaling factors. The input signal matrix is 
\begin{align} 
   \m{X}_k = \begin{bmatrix} \ve{x}_{1,k} & \ve{x}_{2,k} &  \cdots   &  \ve{x}_{M,k} \end{bmatrix} \text{,} 
\end{align} 
with $\ve{x}_{i,k} \in \mathbb{C}^L$ stacking the current and past $L-1$ samples of the $i$th input. In what follows, we proceed with system identification of an unknown complex-valued bilinear system, estimating $\ve{h}$ and $\ve{g}$ via optimum and adaptive filtering techniques.

\subsection*{System identification with noisy input data}
System identification entails finding estimates $\hat{\ve{h}} \in \mathbb{C}^L$ and $\hat{\ve{g}} \in \mathbb{C}^M$ for the unknown vectors $\ve{h}$ and $\ve{g}$, respectively. This can be done by using complex-valued bilinear optimal/adaptive filters.


These bilinear identification methods suffer from a scaling ambiguity since for any nonzero scalar $\nu\in\mathbb{C}$
\begin{align}
   \ve{h}^H \m{X}_k \ve{g} = \left(\nu \ve{h}\right)^H \m{X}_k \ve{g} \frac{1}{\nu^*} \text{.}
\end{align}
Hence, $\ve{h}$ and $\ve{g}$ are identifiable only up to a complex-valued scalar factor. 

For completeness, the bilinear model in \eqref{eq:bilinear-model} can be rewritten as
\begin{align}
   y_k = \ve{f}^T \widetilde{\ve{x}}_k + n_k \text{,} \label{eq:linear-model}
\end{align}
where $\ve{f} = \ve{g} \otimes \ve{h}^* \in \mathbb{C}^{LM}$ and $\widetilde{\ve{x}}_k = \operatorname{vec}\left[\m{X}_k\right] \in \mathbb{C}^{LM}$. Upon this linear form, standard linear optimal/adaptive filters can be applied. Estimating the corresponding coefficient vector $\ve{f}$ eliminates the mentioned ambiguity. However, the number of coefficients to be estimated for this corresponding linear form is $LM$ instead of $L+M$ in the bilinear form. Hence, bilinear methods are often preferred \cite{Meingassner_2026_1, Ghafi_2026_1}.
 
\begin{figure}[!t]
	\centering
	\includegraphics{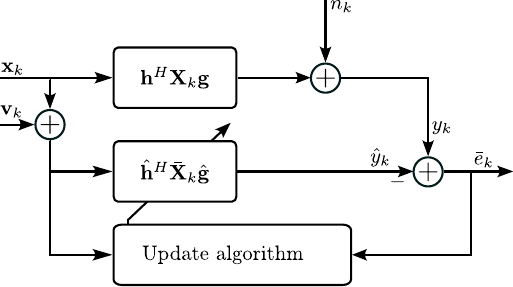}
	\caption{Block diagram of a complex-valued bilinear adaptive filter with noisy input.}
	\label{fig:biasedInput}
\end{figure}

Following \cite{Plaimer_2025_1}, the input signal matrix $\m{X}_k$ and the output $y_k$ are required for the identification process. In practice, however, $\m{X}_k$ is often unavailable and a noisy version $\bar{\m{X}}_k = \m{X}_k + \m{V}_k$ is observed instead (due to e.g., measurement or quantization noise) as illustrated in \autoref{fig:biasedInput}.
For the aforementioned complex-valued \ac{miso} system,
\begin{align}
\m{V}_k = \begin{bmatrix} \ve{v}_{1,k} & \ve{v}_{2,k} &  \cdots   &  \ve{v}_{M,k} \end{bmatrix} \in \mathbb{C}^{L \times M}
\end{align} 
is the input noise matrix, where $\ve{v}_{i,k} \in \mathbb{C}^L$ for $i = 1, \ldots, M$ stacks the current and past $L-1$ samples of the $i$th input noise signal. Hence, an optimum or adaptive filter generates the output signal
\begin{align}
   \hat{y}_k = \hat{\ve{h}}^H \bar{\m{X}}_k \hat{\ve{g}} \text{.} \label{equ:bilinear_filter_output}
\end{align}

The estimates $\hat{\ve{h}}$ and $\hat{\ve{g}}$ are obtained by minimizing a cost function based on the error signal $\bar{e}_k = y_k - \hat{y}_k$. 
However, because the inputs are noisy, the adaptation also suppresses input-noise effects in the error, leading to a bias (i.e., filter coefficients $\hat{\ve{h}}$ and $\hat{\ve{g}}$ become biased toward the axis, as discussed in the next section). To address this bias, the next section derives novel bias-compensated complex-valued bilinear filters.


\section{Bias-Compensated Complex-Valued Bilinear Filters} \label{sec:bias_compensated_complex_valued_bilinear_filters}
The following subsections derive bias-compensated variants of the \ac{cbwf} and the \ac{cblms} filters. Because both require minimizing a real-valued cost function with respect to complex-valued coefficients, we employ the Wirtinger calculus \cite{Wirtinger_1927_1}.
\subsection{Bias-compensated \acs{cbwf1}}
\begin{figure}[!t]
\centering
\includegraphics{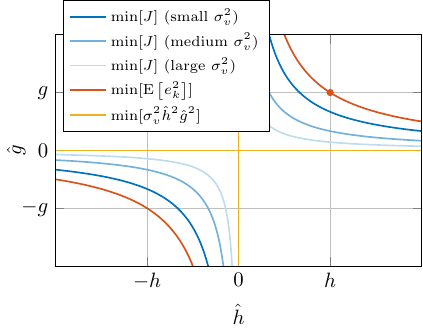}
\caption{Illustrative top-down-view example for the scalar and real-valued cost function \eqref{eq:cost-expanded-scalar}. The blue curves mark the minimum curves of the total cost for increasing input noise variance $\sigma_v^2$ (small, medium, large), the orange curve marks the minimum curve of the first term, and the yellow curves mark the minimum curve of the second term. The true values $h$ and $g$, indicated by the orange dot, lie on the orange curve.}
\label{fig:cost-expanded-scalar-minima}
\end{figure}

To derive the \ac{cbwf}, we minimize the \ac{mse}
\begin{align}
J = \operatorname{E}\left[\bar{e}_k \bar{e}_k^*\right] \in \mathbb{R} \label{eq:cost-standard} \text{.}
\end{align}
By assuming that $\m{X}_k$ and $\m{V}_k$ are statistically independent, this cost function can be written as
\begin{align}
J = \operatorname{E}\left[e_k e_k^*\right] + \operatorname{E}\left[\hat{\ve{h}}^{H}\m{V}_k \hat{\ve{g}} \hat{\ve{g}}^{H}\m{V}_k^{H}\hat{\ve{h}}\right] \label{eq:cost-expanded} \text{,}
\end{align}
where $e_k = y_k - \hat{\ve{h}}^{H}\m{X}_k \hat{\ve{g}}$ denotes the error for noise-free input data.
Minimizing this cost yields biased estimates. This can be illustrated by the real-valued scalar case with $L = M = 1$, where \eqref{eq:cost-expanded} reduces to
\begin{align}
J = \operatorname{E}\left[e_k^2\right] + \sigma_v^2 \hat{h}^2 \hat{g}^2 \text{,} \label{eq:cost-expanded-scalar}
\end{align}
with $\sigma_v^2 = \operatorname{E}\left[v_k^2\right]$ as the input noise variance.
An illustrative example is shown in \autoref{fig:cost-expanded-scalar-minima}. The orange curve indicates the minimum curve of the noise-free term $\operatorname{E}\left[e_k^2\right]$. Clearly, the true values $h$ and $g$ lie on this curve. The term $\sigma_v^2 \hat{h}^2 \hat{g}^2$ is minimized along the axes (yellow curve). In the presence of noise, the adaptive filter minimizes their sum, so the overall minimum curve shifts toward the axes, represented by the blue curves. As $\sigma_v^2$ increases from small to large, this minimum curve moves progressively closer to the axes. Consequently, the standard \ac{cbwf} converges to points on these curves and therefore yields estimates that are not only scaled but also increasingly biased toward the axes as the input noise variance grows.

To overcome this issue, similar to \cite{SungEun_2005_1}, the additional input-noise introduced term in \eqref{eq:cost-expanded} is subtracted from the cost function yielding
\begin{align}
   J_{\text{bc}} = \operatorname{E}\left[\bar{e}_k \bar{e}_k^*\right] - \operatorname{E}\left[\hat{\ve{h}}^H \m{V}_k \hat{\ve{g}} \hat{\ve{g}}^H \m{V}_k^H \hat{\ve{h}}\right] \text{.} \label{eq:cost-bc}
\end{align}
Simple reformulations produces
\begin{align}
   J_{\mathrm{bc}} = \,&\sigma_y^2 - \hat{\ve{h}}^T \m{R}_{\bar{\m{X}}y}^* \hat{\ve{g}}^* - \hat{\ve{h}}^H \m{R}_{\bar{\m{X}}y} \hat{\ve{g}} \label{eq:cost-bc-h-form} \\
   &\qquad + \hat{\ve{h}}^H \left(\bar{\m{R}}_{\ve{g}} - \m{P}_{\ve{g}}\right) \hat{\ve{h}} \nonumber 
\end{align}
and analogously
\begin{align}
   J_{\mathrm{bc}} = \,&\sigma_y^2 - \hat{\ve{h}}^T \m{R}_{\bar{\m{X}}y}^* \hat{\ve{g}}^* - \hat{\ve{h}}^H \m{R}_{\bar{\m{X}}y} \hat{\ve{g}}  \label{eq:cost-bc-g-form} \\
   &\qquad + \hat{\ve{g}}^H \left(\bar{\m{R}}_{\ve{h}} - \m{P}_{\ve{h}}\right) \hat{\ve{g}} \text{,} \nonumber 
\end{align}
where $\sigma_y^2 = \operatorname{E}\left[y_k y_k^*\right] \in \mathbb{C}$, $\m{R}_{\bar{\m{X}}y} = \operatorname{E}\left[\bar{\m{X}}_k y_k^*\right] \in \mathbb{C}^{L \times M}$, $\bar{\m{R}}_{\ve{g}} = \operatorname{E}\left[\bar{\m{X}}_k \hat{\ve{g}} \hat{\ve{g}}^H \bar{\m{X}}_k^H\right] \in \mathbb{C}^{L \times L}$, $\m{P}_{\ve{g}} = \operatorname{E}\left[\m{V}_k \hat{\ve{g}} \hat{\ve{g}}^H \m{V}_k^H\right] \in \mathbb{C}^{L \times L}$, $\bar{\m{R}}_{\ve{h}} = \operatorname{E}\left[\bar{\m{X}}_k^H \hat{\ve{h}} \hat{\ve{h}}^H \bar{\m{X}}_k\right] \in \mathbb{C}^{M \times M}$, and $\m{P}_{\ve{h}} = \operatorname{E}\left[\m{V}_k^H \hat{\ve{h}} \hat{\ve{h}}^H \m{V}_k\right] \in \mathbb{C}^{M \times M}$. 
Similarly to the standard \ac{cbwf} derivation, by equating the Wirtinger gradients of \eqref{eq:cost-bc-h-form} or \eqref{eq:cost-bc-g-form} with respect to $\hat{\ve h}$ and $\hat{\ve g}$ to zero, we obtain
\begin{align}
   \hat{\ve{h}} = \left(\bar{\m{R}}_{\ve{g}} - \m{P}_{\ve{g}}\right)^{-1} \m{R}_{\bar{\m{X}}y} \hat{\ve{g}}^* \label{eq:bwf-h-optimal} 
\end{align}
and
\begin{align}
   \hat{\ve{g}} = \left(\bar{\m{R}}_{\ve{h}} - \m{P}_{\ve{h}}\right)^{-1} \m{R}_{\bar{\m{X}}y}^* \hat{\ve{h}}^* \text{.} \label{eq:bwf-g-optimal}
\end{align}
As derived in \cite{Benesty_2017_1,Plaimer_2025_1}, to overcome the dependency between \eqref{eq:bwf-h-optimal} and \eqref{eq:bwf-g-optimal}, an alternating algorithm can be applied, leading to
\begin{align}
   \hat{\ve{h}}_{n} = \left(\bar{\m{R}}_{\ve{g},n-1} - \m{P}_{\ve{g},n-1}\right)^{-1} \m{R}_{\bar{\m{X}}y} \hat{\ve{g}}_{n-1}^* \label{eq:bwf-h-iter} 
\end{align}
and
\begin{align}
   \hat{\ve{g}}_{n} = \left(\bar{\m{R}}_{\ve{h},n} - \m{P}_{\ve{h},n}\right)^{-1} \m{R}_{\bar{\m{X}}y}^* \hat{\ve{h}}_n^* \text{,} \label{eq:bwf-g-iter}  
\end{align}
with the iteration index $n$, $\bar{\m{R}}_{\ve{g},n-1} = \operatorname{E}\left[\bar{\m{X}}_k \hat{\ve{g}}_{n-1} \hat{\ve{g}}_{n-1}^H \bar{\m{X}}_k^H\right]$, $\m{P}_{\ve{g},n-1} = \\\operatorname{E}\left[\m{V}_k \hat{\ve{g}}_{n-1} \hat{\ve{g}}_{n-1}^H \m{V}_k^H\right]$, $\bar{\m{R}}_{\ve{h},n} = \operatorname{E}\left[\bar{\m{X}}_k^H \hat{\ve{h}}_n \hat{\ve{h}}_n^H \bar{\m{X}}_k\right]$, and $\m{P}_{\ve{h},n} = \operatorname{E}\left[\m{V}_k^H \hat{\ve{h}}_n \hat{\ve{h}}_n^H \m{V}_k\right]$.
Note that $\bar{\m{R}}_{\ve{g},n-1}$, $\m{P}_{\ve{g},n-1}$, $\bar{\m{R}}_{\ve{h},n}$, and $\m{P}_{\ve{h},n}$ can be rewritten as
\begin{align}
   \m{P}_{\ve{h},n} &= \left(\m{I}^{M} \otimes \hat{\ve{h}}_{n} \right)^T \m{R}_{\widetilde{\ve{v}}\widetilde{\ve{v}}}^* \left(\m{I}^{M} \otimes \hat{\ve{h}}_{n}^* \right) \text{,} \label{eq:Ph-kron} \\
   \bar{\m{R}}_{\ve{h},n} &= \left(\m{I}^{M} \otimes \hat{\ve{h}}_{n} \right)^T \left(\m{R}_{\widetilde{\ve{x}}\widetilde{\ve{x}}}^* + \m{R}_{\widetilde{\ve{v}}\widetilde{\ve{v}}}^*\right) \left(\m{I}^{M} \otimes \hat{\ve{h}}_{n}^* \right) \text{,}\label{eq:Rh-kron}\\
   \m{P}_{\ve{g},n-1} &= \left(\hat{\ve{g}}_{n-1} \otimes \m{I}^{L}\right)^T \m{R}_{\widetilde{\ve{v}}\widetilde{\ve{v}}} \left(\hat{\ve{g}}_{n-1}^* \otimes \m{I}^{L}\right) \label{eq:Pg-kron} \\
   \bar{\m{R}}_{\ve{g},n-1} &= \left(\hat{\ve{g}}_{n-1} \otimes \m{I}^{L}\right)^T \left(\m{R}_{\widetilde{\ve{x}}\widetilde{\ve{x}}} + \m{R}_{\widetilde{\ve{v}}\widetilde{\ve{v}}}\right) \left(\hat{\ve{g}}_{n-1}^* \otimes \m{I}^{L}\right) \label{eq:Rg-kron}
\end{align}
with the input covariance matrix $\m{R}_{\widetilde{\ve{x}}\widetilde{\ve{x}}} = \operatorname{E}\left[\widetilde{\ve{x}}_k \widetilde{\ve{x}}_k^H\right]$, the input noise covariance matrix $\m{R}_{\widetilde{\ve{v}}\widetilde{\ve{v}}} = \operatorname{E}\left[\widetilde{\ve{v}}_k \widetilde{\ve{v}}_k^H\right]$, and the corresponding linear noise vector $\widetilde{\ve{v}}_k = \operatorname{vec}\left[\m{V}_k\right]$.

If the underlying system is time-varying or the input statistics change over time, adaptive filters are more suitable. Accordingly, the next subsection derives a sample adaptive and bias-compensated \ac{cblms} filter.

\subsection{Bias-compensated \acs{cblms} filter}
To derive the bias-compensated \ac{cblms} filter, we start from the cost function in \eqref{eq:cost-bc-h-form} or \eqref{eq:cost-bc-g-form}. Using the Wirtinger calculus, the gradients of $J_{\mathrm{bc}}$ with respect to $\hat{\ve{h}}^*$ and $\hat{\ve{g}}^*$, are
\begin{align}
   \frac{\partial{J_{\mathrm{bc}}}}{\partial{\hat{\ve{h}}^*}} = -\m{R}_{\bar{\m{X}}y} \hat{\ve{g}} + \left(\bar{\m{R}}_{\ve{g}} - \m{P}_{\ve{g}}\right) \hat{\ve{h}} \label{eq:grad-h-exact} 
\end{align}
and
\begin{align}
   \frac{\partial{J_{\mathrm{bc}}}}{\partial{\hat{\ve{g}}^*}} = -\m{R}_{\bar{\m{X}}y}^H \hat{\ve{h}} + \left(\bar{\m{R}}_{\ve{h}} - \m{P}_{\ve{h}}\right) \hat{\ve{g}} \text{.} \label{eq:grad-g-exact}
\end{align}
Approximating the statistics $\m{R}_{\bar{\m{X}}y} \approx \bar{\m{X}}_k y_k^*$, $\bar{\m{R}}_{\ve{g}} \approx \bar{\m{X}}_k \hat{\ve{g}} \hat{\ve{g}}^H \bar{\m{X}}_k^H$, and $\bar{\m{R}}_{\ve{h}} \approx \bar{\m{X}}_k^H \hat{\ve{h}} \hat{\ve{h}}^H \bar{\m{X}}_k$ yields 
\begin{align}
   \frac{\partial{J_{\mathrm{bc}}}}{\partial{\hat{\ve{h}}^*}} \approx -\bar{e}_k^* \bar{\m{X}} \hat{\ve{g}} - \m{P}_{\ve{g}} \hat{\ve{h}} \label{eq:grad-h-approx}  
\end{align}
and
\begin{align}
   \frac{\partial{J_{\mathrm{bc}}}}{\partial{\hat{\ve{g}}^*}} \approx -\bar{e}_k \bar{\m{X}}^H \hat{\ve{h}} - \m{P}_{\ve{h}} \hat{\ve{g}} \text{.} \label{eq:grad-g-approx}
\end{align}
Applying the method of steepest descent provides the update equations of the proposed bias-compensated \ac{cblms} filter in the form of
\begin{align}
   \hat{\ve{h}}_{k} = \hat{\ve{h}}_{k-1} + \mu_{\ve{h}} \left(\bar{e}_k^* \bar{\m{X}}_k \hat{\ve{g}}_{k-1} + \m{P}_{\ve{g},k-1} \hat{\ve{h}}_{k-1}\right) \label{eq:blms-h-update}
\end{align}
and
\begin{align}
   \hat{\ve{g}}_{k} = \hat{\ve{g}}_{k-1} + \mu_{\ve{g}} \left(\bar{e}_k \bar{\m{X}}_k^H \hat{\ve{h}}_{k-1} + \m{P}_{\ve{h},k-1} \hat{\ve{g}}_{k-1}\right) \text{,} \label{eq:blms-g-update}
\end{align} 
where $\mu_{\ve{h}} \in \mathbb{R}$ and $\mu_{\ve{g}} \in \mathbb{R}$ denote the step sizes and $\bar{e}_k = y_k - \hat{\ve{h}}_{k-1}^H \bar{\m{X}}_k \hat{\ve{g}}_{k-1}$. Additionally,
\begin{align}
   \m{P}_{\ve{g},k} = \left(\hat{\ve{g}}_{k} \otimes \m{I}^{L}\right)^T \m{R}_{\widetilde{\ve{v}}\widetilde{\ve{v}}} \left(\hat{\ve{g}}_{k}^* \otimes \m{I}^{L}\right) \label{eq:Pg-kron-blms}
\end{align}
and 
\begin{align}
   \m{P}_{\ve{h},k} = \left(\m{I}^{M} \otimes \hat{\ve{h}}_{k} \right)^T \m{R}_{\widetilde{\ve{v}}\widetilde{\ve{v}}}^* \left(\m{I}^{M} \otimes \hat{\ve{h}}_{k}^* \right) \text{.} \label{eq:Ph-kron-blms}
\end{align}

As a last step, the convergence of the filter is analyzed. Assuming $\m{R}_{\widetilde{\ve{v}}\widetilde{\ve{v}}} = \sigma_v^2 \m{I}^{LM}$ and $\m{R}_{\widetilde{\ve{x}}\widetilde{\ve{x}}} = \sigma_x^2 \m{I}^{LM}$, we derive step-size bounds for $\mu_{\ve h}$ and $\mu_{\ve g}$ with an additional constraint on their combination to ensure stability of the proposed bias-compensated \ac{cblms} filter. For that purpose, the error vectors
\begin{align}
	\Delta \ve{h}_{k} = \nu \ve{h} - \hat{\ve{h}}_k  \label{eq:error-vec-h}
\end{align}
and
\begin{align}
	\Delta \ve{g}_{k} = \frac{1}{\nu^*} \ve{g} - \hat{\ve{g}}_k  \text{,}\label{eq:error-vec-g}
\end{align}
are introduced with an arbitrary non-zero scaling factor $\nu \in \mathbb{C}$. These vectors describe the deviation of the estimates $\hat{\ve h}_k$ and $\hat{\ve g}_k$ from the scaled true coefficients $\nu \ve h$ and $\ve g/ \nu^{*}$, respectively. Substituting the update \eqref{eq:blms-h-update} into \eqref{eq:error-vec-h} yields
\begin{align}
   \Delta \ve{h}_{k} = \Delta \ve{h}_{k-1} - \mu_{\ve{h}} \left(\bar{e}_k^* \bar{\m{X}}_k \hat{\ve{g}}_{k-1} + \m{P}_{\ve{g},k-1} \hat{\ve{h}}_{k-1}\right) \text{.} \label{eq:error-h-recursion}  
\end{align}
The second order moment of \eqref{eq:error-h-recursion} can be written as
\begin{align}
	\operatorname{E}\left[||\Delta \ve{h}_{k}||_2^2\right] &= \operatorname{E}\left[||\Delta \ve{h}_{k-1}||_2^2\right] - 2 \mu_{\ve{h}} a_{\ve{h},k} + \mu_{\ve{h}}^2 b_{\ve{h},k} \label{eq:error-norm-h} \text{,}
\end{align}
where 
\begin{align}
   a_{\ve{h},k} = &\operatorname{Re}\left[\operatorname{E}\left[\Delta \ve{h}_{k-1}^H \left(\bar{e}_k^* \bar{\m{X}}_k \hat{\ve{g}}_{k-1} \phantom{.+ \m{P}_{\ve{g},k-1} \hat{\ve{h}}_{k-1}}\right.\right.\right. \nonumber\\
   &\qquad \qquad \qquad \qquad \left.\left.\left.+ \m{P}_{\ve{g},k-1} \hat{\ve{h}}_{k-1}\right)\right]\right] \label{eq:coeff-a-h}
\end{align}
and
\begin{align}
   b_{\ve{h},k} = \operatorname{E}\left[\left(\bar{e}_k^* \bar{\m{X}}_k \hat{\ve{g}}_{k-1} + \m{P}_{\ve{g},k-1} \hat{\ve{h}}_{k-1}\right)^H \right. \nonumber\\
   \left.\left(\bar{e}_k^* \bar{\m{X}}_k \hat{\ve{g}}_{k-1} + \m{P}_{\ve{g},k-1} \hat{\ve{h}}_{k-1}\right)\right] \text{.} \label{eq:coeff-b-h}
\end{align}
With $\m{R}_{\widetilde{\ve{v}}\widetilde{\ve{v}}} = \sigma_v^2 \m{I}^{LM}$ and $\m{R}_{\widetilde{\ve{x}}\widetilde{\ve{x}}} = \sigma_x^2 \m{I}^{LM}$, \eqref{eq:error-norm-h} simplifies to
\begin{align}
   \operatorname{E}\left[||\Delta \ve{h}_{k}||_2^2\right] &= \operatorname{E}\left[||\Delta \ve{h}_{k-1}||_2^2\right] c_{\ve{h},k} + d_{\ve{h},k} \label{eq:error-norm-simplified} \text{,}
\end{align}
where 
\begin{align}
   c_{\ve{h},k} = \, &1 - 2 \mu_{\ve{h}} \sigma_x^2 \operatorname{E}\left[||\hat{\ve{g}}_{k-1}||_2^2\right] + \label{eq:coeff-c-h} \\ 
   &\mu_{\ve{h}}^2 \left(\sigma_x^4 L + \left(\sigma_v^4 + 2 \sigma_v^2 \sigma_x^2\right)\left(L-1\right)\right)\operatorname{E}\left[||\hat{\ve{g}}_{k-1}||_2^2\right]^2 \nonumber
\end{align}
and a corresponding $d_{\ve{h},k}$, which is not given in detail for readability.

It is easy to show that $c_{\ve{h},k} > 0$ for all $\mu_{\ve{h}} > 0$. Furthermore, to avoid a divergent behavior of the proposed bias-compensated \ac{cblms} filter in the mean, $c_{\ve{h},k} < 1$ must hold. This leads to the lower step-size boundary $0 < \mu_{\ve{h}}$ on the one hand and the upper step-size boundary
\begin{align}
  \mu_{\ve{h}} < \frac{2}{\left(\sigma_x^2 L + \left(\frac{\sigma_v^4}{\sigma_x^2} + 2 \sigma_v^2\right)\left(L-1\right)\right)\operatorname{E}\left[||\hat{\ve{g}}_{k-1}||_2^2\right]} \text{.} \label{eq:stepsize-bound}
\end{align}
on the other hand. Similarly, the step-size boundaries for $\mu_{\ve{g}}$ can be derived with
\begin{align}
   c_{\ve{g},k} = \, &1 - 2 \mu_{\ve{g}} \sigma_x^2 \operatorname{E}\left[||\hat{\ve{h}}_{k-1}||_2^2\right] + \label{eq:coeff-c-g} \\ 
   &\mu_{\ve{g}}^2 \left(\sigma_x^4 M + \left(\sigma_v^4 + 2 \sigma_v^2 \sigma_x^2\right)\left(M-1\right)\right)\operatorname{E}\left[||\hat{\ve{h}}_{k-1}||_2^2\right]^2 \nonumber
\end{align}
and a corresponding $d_{\ve{g},k}$ to get the form 
\begin{align}
   \operatorname{E}\left[||\Delta \ve{g}_{k}||_2^2\right] &= \operatorname{E}\left[||\Delta \ve{g}_{k-1}||_2^2\right] c_{\ve{g},k} + d_{\ve{g},k} \label{eq:error-norm-simplified} \text{.}
\end{align}
With that, the step-size is bounded with $0 < \mu_{\ve{g}}$ and 
\begin{align}
    \mu_{\ve{g}} < \frac{2}{\left(\sigma_x^2 M + \left(\frac{\sigma_v^4}{\sigma_x^2} + 2 \sigma_v^2\right)\left(M-1\right)\right)\operatorname{E}[||\hat{\ve{h}}_{k-1}||_2^2]} \text{.} \label{eq:stepsize-bound-g}
\end{align}
To analyze the behavior of the algorithm after convergence on the mean, we investigate these terms for $k \to \infty$ by assuming an appropriate step-size with 
\begin{align}
   \operatorname{E}\left[||\Delta \ve{h}_{\infty}||_2^2\right] &= \lim_{k \to \infty} \operatorname{E}\left[||\Delta \ve{h}_{k}||_2^2\right] \nonumber \\
    &= \lim_{k \to \infty} \frac{d_{\ve{h},k}}{1 - c_{\ve{h},k}} = \frac{d_{\ve{h},\infty}}{1 - c_{\ve{h},\infty}} \label{eq:error-norm-h-infty}
\end{align}
and
\begin{align}
    \operatorname{E}\left[||\Delta \ve{g}_{\infty}||_2^2\right] &= \lim_{k \to \infty} \operatorname{E}\left[||\Delta \ve{g}_{k}||_2^2\right] \nonumber \\
    &= \lim_{k \to \infty} \frac{d_{\ve{g},k}}{1 - c_{\ve{g},k}} = \frac{d_{\ve{g},\infty}}{1 - c_{\ve{g},\infty}} \text{.} \label{eq:error-norm-g-infty}
\end{align}
The coefficients $c_{\ve{h},\infty}$ and $c_{\ve{g},\infty}$ denote the steady-state limits of $c_{\ve{h},k}$ in \eqref{eq:coeff-c-h} and $c_{\ve{g},k}$ in \eqref{eq:coeff-c-g}, obtained by replacing $\operatorname{E}\left[||\hat{\ve{g}}_{k-1}||_2^2\right]$ and $\operatorname{E}\left[||\hat{\ve{h}}_{k-1}||_2^2\right]$ with their steady-state values $\operatorname{E}\left[||\hat{\ve{g}}_{\infty}||_2^2\right]$ and $\operatorname{E}\left[||\hat{\ve{h}}_{\infty}||_2^2\right]$. Analogously, the steady-state terms $d_{\ve{g},\infty}$ and $d_{\ve{h},\infty}$ are given in \eqref{eq:coeff-d-g-infty} and \eqref{eq:coeff-d-h-infty}, respectively.


\begin{figure*}[!t]
   \centering
   \small
   \begin{align}
      d_{\ve{g},\infty} &= \underbrace{\mu_{\ve{g}}^2 M \sigma_x^2 \left(\sigma_x^2 + \sigma_v^2\right) ||\ve{g}||_2^2 ||\ve{h}||_2^2}_{\displaystyle \kappa_{\ve{g}}}~\operatorname{E}\left[||\Delta \ve{h}_{\infty}||_2^2\right] \nonumber \\
      &+ \underbrace{\mu_{\ve{g}}^2 |\nu|^2 ||\ve{h}||_2^2 \left(M  \left(\sigma_x^2 + \sigma_v^2\right) \left(\sigma_n^2 + \sigma_v^2 ||\ve{g}||_2^2 ||\ve{h}||_2^2 \right) - \sigma_v^4 ||\ve{g}||_2^2 ||\ve{h}||_2^2 \right)}_{\displaystyle \gamma_{\ve{g}}} \label{eq:coeff-d-g-infty}
   \end{align}
   \begin{align}
      d_{\ve{h},\infty} &= \underbrace{\mu_{\ve{h}}^2 L \sigma_x^2 \left(\sigma_x^2 + \sigma_v^2\right) ||\ve{g}||_2^2 ||\ve{h}||_2^2}_{\displaystyle \kappa_{\ve{h}}}~\operatorname{E}\left[||\Delta \ve{g}_{\infty}||_2^2\right] \nonumber \\
      &+ \underbrace{\mu_{\ve{h}}^2 \frac{1}{|\nu|^2} ||\ve{g}||_2^2 \left(L  \left(\sigma_x^2 + \sigma_v^2\right) \left(\sigma_n^2 + \sigma_v^2 ||\ve{g}||_2^2 ||\ve{h}||_2^2 \right) - \sigma_v^4 ||\ve{g}||_2^2 ||\ve{h}||_2^2 \right)}_{\displaystyle \gamma_{\ve{h}}} \label{eq:coeff-d-h-infty}
   \end{align}
   \rule{\textwidth}{0.4pt}
\end{figure*}

With that, the linear system
\begin{align}
   \left(1 - c_{\ve{h},\infty}\right) \operatorname{E}\left[||\Delta \ve{h}_{\infty}||_2^2\right] - \kappa_{\ve{h}} \operatorname{E}\left[||\Delta \ve{g}_{\infty}||_2^2\right] = \gamma_{\ve{h}} \label{eq:linsys-h}
\end{align}
and
\begin{align}
   \left(1 - c_{\ve{g},\infty}\right) \operatorname{E}\left[||\Delta \ve{g}_{\infty}||_2^2\right] - \kappa_{\ve{g}} \operatorname{E}\left[||\Delta \ve{h}_{\infty}||_2^2\right] = \gamma_{\ve{g}} \text{,} \label{eq:linsys-g}
\end{align}
is obtained by inserting \eqref{eq:coeff-d-g-infty} and \eqref{eq:coeff-d-h-infty} into \eqref{eq:error-norm-g-infty} and \eqref{eq:error-norm-h-infty}. Subsequently, this can be solved to
\begin{align}
    \operatorname{E}\left[||\Delta \ve{h}_{\infty}||_2^2\right] = \frac{\gamma_{\ve{h}} \left(1 - c_{\ve{g},\infty}\right) + \kappa_{\ve{h}} \gamma_{\ve{g}}}{\Delta}
\end{align}
and 
\begin{align}
    \operatorname{E}\left[||\Delta \ve{g}_{\infty}||_2^2\right] = \frac{\gamma_{\ve{g}} \left(1 - c_{\ve{h},\infty}\right) + \kappa_{\ve{g}} \gamma_{\ve{h}}}{\Delta}
\end{align}
with
\begin{align}
   \Delta = \left(1 - c_{\ve{h},\infty}\right)\left(1 - c_{\ve{g},\infty}\right) - \kappa_{\ve{h}} \kappa_{\ve{g}} \text{.} \label{eq:Delta-def}
\end{align}
From that it is clear that $\Delta > 0$ should hold, which provides an additional restriction for the step-size boundaries.

Note that if the input noise is negligible, i.e., $\sigma_v^2 \approx 0$, the step-size boundaries of the proposed bias-compensated \ac{cblms} filter are identical to the step-size boundaries of the standard \ac{cblms} filter in \cite{Plaimer_2025_1}.

\section{Estimation of Input Noise Variance}\label{sec:estimation_of_noise_variance}
All the bias-compensated filters discussed above have one aspect in common. They require knowledge of the noise statistics, which is not always available in practice. To address this, a strategy from \cite{SungEun_2005_1} is adapted in the following to estimate the unknown input-noise variance from the observed data. As derived below, this estimation additionally requires knowledge of the output-noise variance $\sigma_n^2$, which, unlike $\sigma_v^2$, is assumed known and is not estimated as part of this derivation. Its estimation is well studied in the literature and is heavily application-specific, may be obtained trivially by disabling the input and measuring the resulting output power, or by more sophisticated methods \cite{Schuster_2019_1,Söderström_2007_1}. After convergence of the filters, the cost function in \eqref{eq:cost-bc} equals the output-noise variance, i.e., $J_{\text{bc}} = \sigma_n^2$, yielding
\begin{align}
   \sigma_n^2 = \operatorname{E}\!\left[\bar{e}_k \bar{e}_k^*\right] - \operatorname{E}\!\left[\hat{\ve{h}}^H \m{V}_k \hat{\ve{g}}\, \hat{\ve{g}}^H \m{V}_k^H \hat{\ve{h}}\right] \text{.} \label{eq:estimateVar_costFunction}
\end{align}
Using $\hat{\ve{h}}^H \m{V}_k \hat{\ve{g}} = \hat{\ve{f}}^T \widetilde{\ve{v}}_k$, with $\hat{\ve{f}} = \hat{\ve{g}} \otimes \hat{\ve{h}}^* \in \mathbb{C}^{LM}$ and $\widetilde{\ve{v}}_k = \operatorname{vec}[\m{V}_k]$, the right-hand side of \eqref{eq:estimateVar_costFunction} can be rewritten as 
\begin{align}
   \operatorname{E}\!\left[\hat{\ve{h}}^H \m{V}_k \hat{\ve{g}}\, \hat{\ve{g}}^H \m{V}_k^H \hat{\ve{h}}\right]
   &= \hat{\ve{f}}^T\, \operatorname{E}\!\left[\widetilde{\ve{v}}_k \widetilde{\ve{v}}_k^H\right] \hat{\ve{f}}^* \\
   &= \hat{\ve{f}}^T\, \m{R}_{\widetilde{\ve{v}}\widetilde{\ve{v}}}\, \hat{\ve{f}}^* \text{.} \nonumber
\end{align}
With $\m{R}_{\widetilde{\ve{v}}\widetilde{\ve{v}}} = \sigma_v^2 \m{I}^{LM}$ and by applying the Kronecker norm identity $\|\hat{\ve{g}} \otimes \hat{\ve{h}}^*\|_2^2 = \|\hat{\ve{g}}\|_2^2\, \|\hat{\ve{h}}\|_2^2$, one obtains
\begin{align}
   \hat{\ve{f}}^T\, \m{R}_{\widetilde{\ve{v}}\widetilde{\ve{v}}}\, \hat{\ve{f}}^*
   = \sigma_v^2\, \|\hat{\ve{f}}\|_2^2
   = \sigma_v^2 \|\hat{\ve{g}}\|_2^2\, \|\hat{\ve{h}}\|_2^2 \text{.} \label{eq:rh}
\end{align}
Replacing $\operatorname{E}[\bar{e}_k \bar{e}_k^*]$ with the instantaneous estimate $\bar{e}_k \bar{e}_k^*$, and $\hat{\ve{h}}$, $\hat{\ve{g}}$ with $\hat{\ve{h}}_k$, $\hat{\ve{g}}_k$, substituting \eqref{eq:rh} into \eqref{eq:estimateVar_costFunction}, and solving for $\sigma_v^2$ yields the estimate
\begin{align}
   \hat{\sigma}_{v,k}^2 = \frac{\bar{e}_k \bar{e}_k^* - \sigma_n^2}{\|\hat{\ve{g}}_k\|_2^2\, \|\hat{\ve{h}}_k\|_2^2 + \delta} \text{,} \label{eq:sigma-estimator}
\end{align}
with an additional small positive constant $\delta$ to avoid a division by zero. In practice, the instantaneous estimate $\bar{e}_k \bar{e}_k^*$ is noisy, which motivates the use of a more robust variant. The estimator from \cite{SungEun_2005_1} replaces the numerator and denominator with smoothed estimates using an exponential moving average with a smoothing factor $\beta \in (0,1)$. The initial values are set to $\Phi_{e^2,0} = \sigma_n^2$ and $\Phi_{f^2,0} \geq 0$. The smoothed estimator is subsequently given by
\begin{align}
\hat{\sigma}_{v,k}^2 = \frac{\Phi_{e^2,k}  - \sigma_n^2}{ \Phi_{f^2,k} + \delta} \label{eq:sigma-psi}
\end{align}
with
\begin{align}
   \Phi_{e^2,k} &= (1-\beta)\,\Phi_{e^2,k-1} + \beta\,\bar{e}_k \bar{e}_k^* \text{,} \\
   \Phi_{f^2,k} &= (1-\beta)\,\Phi_{f^2,k-1} + \beta\|\hat{\ve{g}}_k\|_2^2\, \|\hat{\ve{h}}_k\|_2^2 \text{.}
\end{align}

The derived filter might yield a divergent adaptation for a wrongly estimated input noise variance as stated in \cite{SungEun_2005_1}. To tackle this issue, an upper bound for $\hat{\sigma}_{s,k}^2$ is derived starting with the condition
\begin{align}
   \operatorname{E}\left[ \tilde{\bar{\ve{x}}}_k^{\operatorname{T}} \tilde{\bar{\ve{x}}}_k \right] = \operatorname{E}\left[ \widetilde{\ve{x}}_k^{\operatorname{T}} \widetilde{\ve{x}}_k \right] + \operatorname{E}\left[ \widetilde{\ve{v}}_k^{\operatorname{T}} \widetilde{\ve{v}}_k \right] \text{.} \label{eq:power-balance}
\end{align}
From \eqref{eq:power-balance}, the upper bound is found by
\begin{align}
   \operatorname{E}\left[ \tilde{\bar{\ve{x}}}_k^{\operatorname{T}} \tilde{\bar{\ve{x}}}_k \right] - \operatorname{E}\left[ \widetilde{\ve{v}}_k^{\operatorname{T}} \widetilde{\ve{v}}_k \right] &\geq 0 \text{,} \nonumber \\
   \|\tilde{\bar{\ve{x}}}_k\|_2^2 - LM\, \sigma_{v}^2 &\gtrsim 0 \text{,}
\end{align}
with $\operatorname{E}\left[ \tilde{\bar{\ve{x}}}_k^{\operatorname{T}} \tilde{\bar{\ve{x}}}_k \right] \approx \|\tilde{\bar{\ve{x}}}_k\|_2^2$. This is now used as an upper bound for the estimated input noise variance $\hat{\sigma}_{v,k}^2$ as
\begin{align}
       \hat{\sigma}_{v,k}^2 = \min\left[\frac{\Phi_{e^2,k}  - \sigma_n^2}{ \Phi_{f^2,k} + \delta}, \frac{|| \tilde{\bar{\ve{x}}}_k||^2_2}{LM}\right]  \text{.} \label{eq:sigma-upper-bound}
\end{align}
Analogously this input noise variance estimation can be applied to the \ac{cbwf}, however, using the iteration $n$ instead of the sample index $k$.

\section{Simulations}\label{sec:simulations}
To evaluate the performance of the proposed bias-compensated complex-valued bilinear filters the complex-valued bilinear system model in \eqref{eq:bilinear-model} is considered with coefficient vector lengths $L = 30$ and $M = 20$. Each entry of the true vector $\ve{h}$ is modelled by $h_i \sim \mathcal{CN}(0,\, \sigma_i^2)$. The per-tap variance follows an exponential decay profile $\sigma_i^2 = \sigma_0^2\, e^{-0.7(i-1)}$ with $\sigma_0^2 = 1 - e^{-0.7}$ and $i = 1 \dots L$. The vector $\ve{g}$ is drawn as a complex-valued circular Gaussian random vector and normalized to unit norm $\|\ve{g}\| = 1$. The input signal matrix $\m{X}_k$ for a \ac{miso} system is modelled by independent and identically distributed complex-valued circular Gaussian samples with unit variance and zero mean. The corresponding input noise matrix $\m{V}_k$ is assumed to be independent and identically distributed complex-valued circular Gaussian with variance $\sigma_v^2$, and the circular Gaussian output noise $n_k$ has the variance $\sigma_n^2$. In this work, simulations are conducted with an output noise variance of $\sigma_n^2 = 0.2$ and $\sigma_n^2= 1$, to illustrate the behavior for different variances.

For the sample-adaptive bilinear filters, the step sizes are set to $\mu_{\ve{h}} = \mu_{\ve{g}} = 5 \times 10^{-5}$ . The filters are initialized with independent complex-valued circular Gaussian random vectors, where $\hat{\ve{h}}_0 \sim \mathcal{CN}\!\left(\ve{0},\, \frac{1}{L}\m{I}\right)$ and $\hat{\ve{g}}_0 \sim \mathcal{CN}\!\left(\ve{0},\, \frac{1}{M}\m{I}\right)$. For the input noise variance estimation, the smoothing factor is set to $\beta = 1 \times 10^{-3}$ and $\delta = 1 \times 10^{-3}$. 

The proposed filters are compared to four state-of-the-art bias-compensated algorithms originally developed for widely linear or linear systems, namely the \ac{lceilms} \cite{Zhang_2019_1}, the \ac{wlceimccc} \cite{Dong_2020_1}, the \ac{vmmtcc} \cite{Cai_2025_1}, and the \ac{acgdtls} \cite{Zhang_2025}. As described above, the underlying complex-valued bilinear model in \eqref{eq:bilinear-model} can be written in a corresponding linear form $\ve{f} = \ve{g} \otimes \ve{h}^*$ and all mentioned state-of-the-art filters are applied to this corresponding linear model. All four reference algorithms require known input and output noise variances $\sigma_v^2$ and $\sigma_n^2$, whereas the proposed bias-compensated filters additionally provide results using estimated input noise variances. 
\begin{figure}[t]
	\centering
	\includegraphics{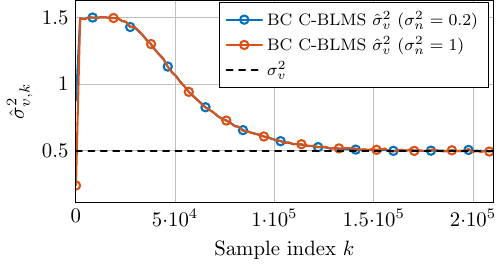}
	\caption{Estimated input noise variance $\hat{\sigma}_{v,k}^2$ using the bias-compensated \ac{cblms} over sample index $k$ for $\sigma_v^2 = 0.5$ and different output noise variances $\sigma_{n}^2$. Here, \acs{bc} abbreviates bias-compensated for readability.}
	\label{fig:sigma_est_conv}
\end{figure}
\begin{figure}[!t]
	\centering
	\includegraphics{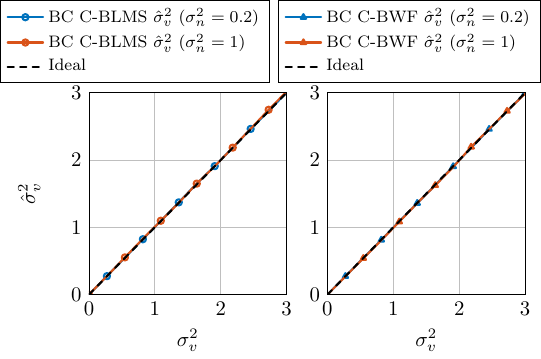}
	\caption{Estimated input noise variance $\hat{\sigma}_{v}^2$ after convergenceversus the true input noise variance $\sigma_v^2$ for the bias-compensated \ac{cblms} (left) and bias-compensated \ac{cbwf} (right) with different output noise variances $\sigma_{n}^2$, respectively. Here, \acs{bc} abbreviates bias-compensated for readability.}
	\label{fig:sigma_est_mc}
	\label{fig:sigma_est_bwf_mc}
\end{figure}
\begin{figure}[t]
	\centering
	\includegraphics{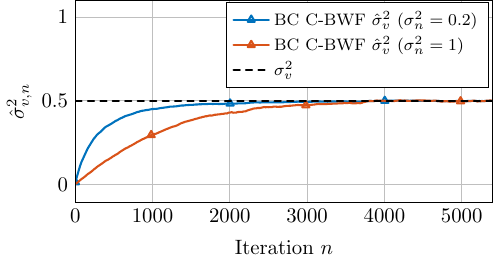}
	\caption{Estimated input noise variance $\hat{\sigma}_{v,n}^2$ using the bias-compensated \ac{cbwf} over iteration $n$ for $\sigma_v^2 = 0.5$ and different output noise variances $\sigma_{n}^2$. Here, \acs{bc} abbreviates bias-compensated for readability.}
	\label{fig:sigma_est_bwf_conv}
\end{figure}
\subsection{Estimated Noise Variance}
Before investigating the performance of the specific bilinear filters, the input noise variance estimation is evaluated as the ensemble mean over $50$ simulation runs with $\sigma_v^2 = 0.5$. \autoref{fig:sigma_est_conv} shows the estimated noise variance $\hat{\sigma}_{v,k}^2$ of the bias-compensated \ac{cblms} over the sample index $k$, and \autoref{fig:sigma_est_bwf_conv} shows the corresponding estimate $\hat{\sigma}_{v,n}^2$ of the bias-compensated \ac{cbwf} over the iteration index $n$, each for the two output noise variances $\sigma_n^2 = 0.2$ and $\sigma_n^2 = 1$. In both cases, the estimates converge closely to the true input noise variance $\sigma_v^2 = 0.5$ (dashed line).

\autoref{fig:sigma_est_mc} shows the estimated noise variance after convergence, averaged over $100$ Monte Carlo runs, as a function of the true input noise variance $\sigma_v^2$ for the bias-compensated \ac{cblms} (left) and the bias-compensated \ac{cbwf} (right). The diagonal dashed line (ideal) represents a perfect noise variance estimation across the full range of input noise levels. For both filters and both output noise variances $\sigma_n^2 = 0.2$ and $\sigma_n^2 = 1$, the estimated noise variance closely tracks the true value over the considered range of $\sigma_v^2$, confirming that the proposed estimator is reliable for different output noise levels.

\subsection{Normalized Misalignment}

To evaluate the coefficient estimation quality, we use the \ac{nm} defined as $\text{NM}\left(\hat{\ve{f}}_k\right) = \|\hat{\ve{f}}_k - \ve{f}\|_2^2/\|\ve{f}\|_2^2$, where $\hat{\ve{f}}_k = \hat{\ve{g}}_k \otimes \hat{\ve{h}}_k^*$ is the linear coefficient vector at sample index $k$ for the sample adaptive filter and analogously at the iteration $n$ for the Wiener filters.

\begin{figure}[!t]
	\centering
	\includegraphics{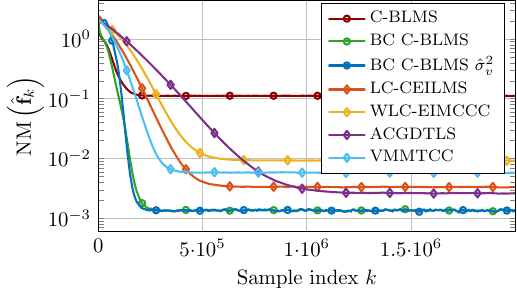}
	\caption{Normalized misalignment over sample index $k$ averaged over 50 simulation runs with $\sigma_v^2 = 0.5$ and $\sigma_n^2 = 0.2$, comparing the standard version and the bias-compensated \ac{cblms} filters with the state of the art. Here, \acs{bc} abbreviates bias-compensated for readability.}
	\label{fig:convergence02}
\end{figure}

\begin{figure}[!t]
	\centering
	\includegraphics{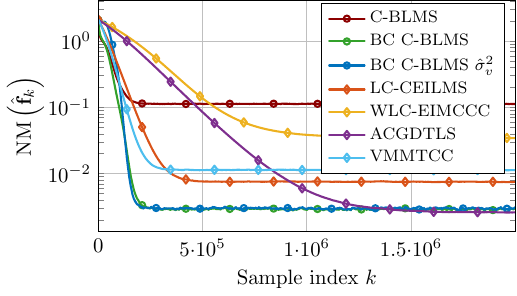}
	\caption{Normalized misalignment over sample index $k$ averaged over 50 simulation runs with $\sigma_v^2 = 0.5$ and $\sigma_n^2 = 1$, comparing the standard version and the bias-compensated \ac{cblms} filters with the state of the art. Here, \acs{bc} abbreviates bias-compensated for readability.}
	\label{fig:convergence1}
\end{figure}

\autoref{fig:convergence02} shows the \ac{nm} over the sample index $k$, as the ensemble mean over $50$ simulation runs with $\sigma_v^2 = 0.5$ and $\sigma_n^2 = 0.2$, comparing the standard \ac{cblms} filter, the proposed bias-compensated \ac{cblms} filter (with true and estimated input noise variance), and the four state-of-the-art reference algorithms (which use the corresponding linear model \eqref{eq:linear-model}). The standard \ac{cblms} filter converges to a biased solution with a high steady-state misalignment, whereas the proposed bias-compensated \ac{cblms} filters achieve a significantly lower steady-state misalignment. The state-of-the-art algorithms, which require the true noise variances $\sigma_v^2$ and $\sigma_n^2$, are clearly outperformed by the proposed filters. The corresponding results for $\sigma_n^2 = 1$ are shown in \autoref{fig:convergence1} and exhibit a similar behavior, with one notable exception. For this larger output noise variance, the \ac{acgdtls} achieves a slightly lower steady-state misalignment than the proposed bias-compensated \ac{cblms} filters for $\sigma_v^2 = 0.5$. However, as shown next, this advantage does not persist for larger input noise variances.

The \ac{cbwf} and bias-compensated \ac{cbwf} are not included in \autoref{fig:convergence02} and \autoref{fig:convergence1}, since they are indexed by iteration $n$ rather than by sample index $k$. They converge within a few iterations, several orders of magnitude faster than the \ac{cblms} filters previously discussed. The only exception is the bias-compensated \ac{cbwf} using the estimated input noise variance, which requires a short additional adaptation period until the noise variance estimate itself has settled. However, the steady-state behavior is investigated in the following.

\begin{figure}[!t]
	\centering
	\includegraphics{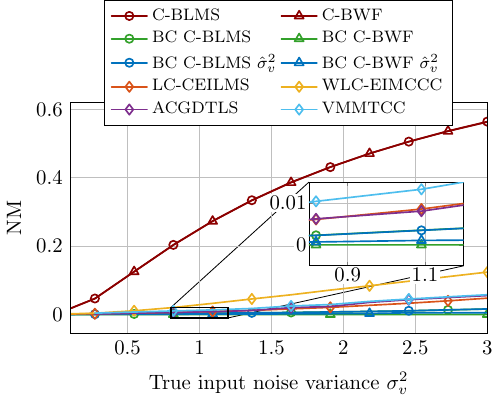}
	\caption{Normalized misalignment after convergence over the true input noise variance $\sigma_v^2$. The output noise variance is set to $\sigma_n^2 = 0.2$. Here, \acs{bc} abbreviates bias-compensated for readability.}
	\label{fig:NM_mc02}
\end{figure}

\subsection{Steady-state performance}
To investigate the steady-state performance as a function of the input noise variance, Monte Carlo simulations are conducted. The input noise variance $\sigma_v^2$ is varied from $0$ to $3$, and for each noise level, $N_{\text{sim}} = 100$ independent runs are performed, each with $N_{\text{samples}} = 2 \times 10^6$ samples and the results are averaged over all runs.

\begin{figure}[!t]
	\centering
	\includegraphics{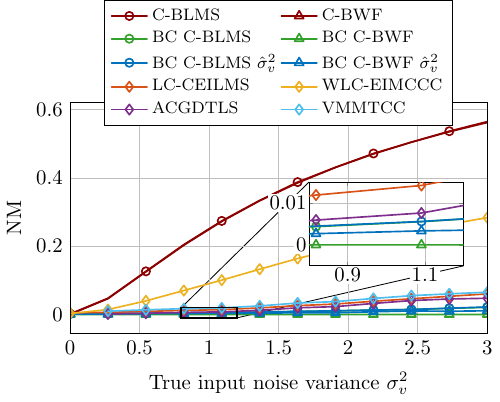}
	\caption{Normalized misalignment after convergence over the true input noise variance $\sigma_v^2$. The output noise variance is set to $\sigma_n^2 = 1$. Here, \acs{bc} abbreviates bias-compensated for readability.}
	\label{fig:NM_mc1}
\end{figure}

\autoref{fig:NM_mc02} shows the \ac{nm} after convergence as a function of $\sigma_v^2$ with an output noise variance of $\sigma_n^2 = 0.2$. For the uncompensated \ac{cblms} filter and the \ac{cbwf}, the \ac{nm} increases significantly when increasing input noise variance, demonstrating the bias effect. The proposed bias-compensated \ac{cblms} and \ac{cbwf} filters maintain a substantially lower \ac{nm} across the whole range of $\sigma_v^2$, both when the true input noise variance is used and when it is estimated. The state-of-the-art reference algorithms are clearly outperformed by the proposed filters. The corresponding results for $\sigma_n^2 = 1$ are shown in \autoref{fig:NM_mc1} illustrating a similar behavior, except for the \ac{acgdtls}. It achieves a slightly lower steady-state misalignment than the proposed bias-compensated \ac{cblms} filters at small input noise variances, however, as $\sigma_v^2$ increases the proposed bias-compensated \ac{cblms} filters also outperform the \ac{acgdtls}.

\section{Conclusion}\label{sec:conclusion}
In this work, we addressed the bias that occurs in complex-valued bilinear filters when the input signal is corrupted by noise, e.g., due to measurement or quantization errors. To this end, we derived a bias-compensated \ac{cbwf} and a bias-compensated \ac{cblms} filter based on an adapted cost function that accounts for the input-noise statistics. For the bias-compensated \ac{cblms} filter, we additionally provided a convergence analysis, yielding step-size boundaries. Because the proposed filters require knowledge of the input-noise variance, we also investigated a method for estimating this quantity from the observed data. Simulation results on a complex-valued \ac{miso} system showed that the proposed bias-compensated filters substantially outperform both their standard counterparts, and state-of-the-art bias-compensated algorithms, across a wide range of input-noise levels.

\bibliography{bibliography.bib}
\end{document}